\documentclass[prb,twocolumn,showpacs,floatfix]{revtex4}
\usepackage{pdfsync}
\usepackage{times}
\usepackage{bm}
\usepackage[dvips]{graphicx}
\usepackage{amsmath}
\usepackage{natbib}
\usepackage{color}
\usepackage{epstopdf}
\usepackage{gensymb}

\begin{document}

\title{Quantum Transport in Chemically-modified Two-Dimensional Graphene: From Minimal Conductivity to Anderson Localization}

\author{N. Leconte$^{1}$, A. Lherbier$^{1}$, F. Varchon$^{1}$, P. Ordejon$^{2}$, S. Roche$^{3,4}$, J.-C. Charlier$^{1}$}
\affiliation{
$^1$Universit\'e catholique de Louvain, Institut de la Mati\`ere Condens\'ee et des Nanosciences (IMCN), NAPS-ETSF, Chemin des Etoiles 8, B-1348 Louvain-la-Neuve, Belgium\\
$^2$Centre de Investigaci\'o en Nanoci\`encia i Nanotecnologia, CIN2 (CSIC-ICN), Campus de la UAB, 08193 Bellaterra (Barcelona), Spain\\ 
$^3$CIN2 (ICN-CSIC) and Universitat Autonoma de Barcelona, Catalan Institute of Nanotechnology, Campus de la UAB, 08193 Bellaterra (Barcelona), Spain\\
$^4$ICREA, Institució Catalana de Recerca i Estudis Avancats, 08010 Barcelona, Spain
}

\date{\today}

\begin{abstract} 
  An efficient computational methodology is used to explore charge transport properties in chemically-modified (and randomly disordered) graphene-based materials. The Hamiltonians of various complex forms of graphene are constructed using tight-binding models enriched by first-principles calculations. These atomistic models are further implemented into a real-space order-N Kubo-Greenwood approach, giving access to the main transport length scales (mean free paths, localization lengths) as a function of defect density and charge carrier energy. An extensive investigation is performed for epoxide impurities with specific discussions on both the existence of a minimum semi-classical conductivity and a crossover between weak to strong localization regime. The 2D generalization of the Thouless relationship linking transport length scales is here illustrated based on a realistic disorder model. 
\end{abstract} 


\maketitle

\section{Introduction}
Ever since graphene was experimentally synthesized in 2004\cite{Novoselov1}, interest in its promising conduction properties has increased considerably\cite{Novoselov2, Novoselov3,Novoselov4}. Owing to its two-dimensionality and reported large charge mobility, monolayer graphene has been initially envisioned as a genuine candidate to replace silicon in nano-electronics\cite{Novoselov4,carbonElectronics1, carbonElectronics2}. But despite its realistic potential in high-frequency device applications\cite{highFrequency1, highFrequency2}, the absence of a substantial band-gap hinders its use for replacing silicon MOSFET devices in logic applications\cite{highFrequency3}. 

Various solutions have already been proposed to overcome this hurdle such as opening a wide band-gap, using quantum confinement in ribbons\cite{confin1, confin2, confin3} or using chemical oxidation or hydrogenation to break the symmetry of the graphene lattice\cite{OXGRAPHANE1, OXGRAPHANE2, OXGRAPHANE3, OXGRAPHANE4, OXGRAPHANE5, OXGRAPHANE6, OXGRAPHANE7, OXGRAPHANE8, OXGRAPHANE9, OXGRAPHANE10}. Both of these methods have however been demonstrated to be far too invasive\cite{highFrequency3,highFrequency4,highFrequency5}, generating a large quantity of defects and damaging the otherwise Dirac-like properties of electronic excitations. Other more seducing proposals include the use of a laser field in the mid-infrared range which can induce tunable band gaps\cite{tuneBandGap1}, electric-field assisted gap opening in bilayers\cite{McCann} or chemical doping which in certain conditions allow to engineer controlled mobility gaps as large as $1$ eV\cite{Biel1, Biel2}. 

In all cases, the precise understanding of the impact of disorder on electronic and (charge, spin and phonon) transport properties of graphene appears of paramount importance. Disorder in graphene exhibits many different flavors from structural defects to adsorbed impurities, reconstructed edges or long range Coulomb scatterers trapped in the graphene substrate (oxide layer). To date, the detailed relationship between microscopic complexity of disorder features and the onset of graphene unique transport properties remains elusive. This is particularly debated in relation with the so-called Klein tunneling mechanism\cite{KleinTunnel} and the weak anti-localization phenomenon which are both manifestations of pseudospin effects\cite{WALTHEO3, SL1, SL2, SL3, Ortmann}.

Disorder first comes as a source of elastic scattering which limits the mean free path in a
way which strongly depends on the disorder potential characteristics. The energy
dependence of the mean free path and associated semi-classical transport quantities such
as the Drude conductivity and the charge mobility can be indeed connected to the short
or long range nature of the scattering potential\cite{dasSarma}. Beyond the occurrence of a diffusive
regime, quantum interferences contribute significantly to the transport features at
sufficiently low temperatures. In addition to the conventional weak localization
phenomenon\cite{Ramakrishnan, Akkermans}, crossovers from weak localization to weak anti-localization have
been predicted and experimentally observed\cite{WALTHEO1,WALTHEO2,WALTHEO3,WALTHEO4,WALTHEO5,WALTHEO6,WALEXP1,WALEXP2,SL1,SL2,Ortmann}. Pseudospin-related
quantum interferences are however maintained provided disorder does not break all
underlying symmetries. This is not the case in presence of chemical defects which
damage the $sp^{2}$ lattice symmetry. Such 
stronger disturbances of graphene structure maximize localization effects, eventually
turning the material to a two-dimensional insulator (Anderson localization). If Anderson
localization has been highly debated and controversial for long-range disorder\cite{SL4,SL5,SL6,SL7,SL8,SL9,SL10,SL11}, its
relevance for strongly damaged graphene is now well documented both theoretically
and experimentally\cite{Leconte,hydrogen,fluor,WeiLi}. A recent theoretical study has however related the existence of
a robust metallic state in presence of local magnetic ordering for partly hydrogenated graphene\cite{Leconte1,Leconte2} pinpointing possible subtleties between correlated impurity distribution and transport features. 

The main objective of this paper is to illustrate how an insulating regime can be tuned by intrusive functionalization of a graphene sheet caused by oxygen atoms bound in the epoxy position. Epoxide defects are for instance incorporated on graphene after ozone treatment\cite{Lee}. These epoxy impurities have a drastically different impact on resonant energy peaks in the vicinity of the Dirac point when compared to single impurities\cite{Wehling}. 

To address this objective, the oxygen in epoxy position is studied by means of accurate {\it ab initio} techniques. This model allows us, on the one hand, to prove that the oxygen epoxy bonding lies in between a pure $sp^3$-like covalent bond and an ionic bond, 
and on the other hand, to supply a suitable tight-binding (TB) model for further studies in very large scale systems. Using this TB model, the Kubo-Greenwood formalism is implemented in real space to obtain meaningful transport length scales and conduction properties. Several quantities such as the mean free path, the semi-classical conductivity and the localization length are analyzed in depth. The ongoing debate concerning long and short-range scattering behavior is also briefly commented in light of our results. The crossover to the strongly localized regime is then investigated. Finally, conventional scaling laws are tested on our model in this localization regime.

\section{Epoxy defects}
  
V.V. Cheianov {\it et al.} [\onlinecite{Cheianov}] demonstrated the tendency of epoxy-bound adatoms to form spatially correlated states. The interaction between epoxy groups is mediated by the
conduction electrons, similar to the Ruderman-Kittel-Kasuya-Yosida (RKKY) interaction which correlates magnetic impurities\cite{Abanin}. These ordered states only exist for low impurity densities and disappear at a critical
temperature $T_c$. For high concentrations of epoxy groups due to oxidation\cite{Staudenmaier} of graphene, J.L. Li {\it et al.} [\onlinecite{Li}] argue that two epoxy groups attaching on the
opposite ends of a carbon hexagon create more open rings inducing cracks along neighboring rings. Similarly, more recently, S. Fabris {\it et al.}[\onlinecite{Sun}] present a mechanism giving rise to
more complex crack propagation. However, crack propagation has, up to our knowledge, only been reported under strong reduction and oxidation treatments. Furthermore, an aqueous environment seems to be mandatory. H.J. Xiang {\it et al.} [\onlinecite{GongXinGao}] also report on these unzipped chains caused by epoxy groups thus inducing lower energy conformations. Their study is however limited to
concentrations above $25$ \% of epoxy density.

The model investigated in the present paper takes advantage of this literature while limiting its complexity to avoid any loss of generality for the simulations of different moderate concentrations
(epoxy density ranging from $0.01$ \% to $5$ \%). Consequently, epoxy groups are assumed to be randomly distributed over the graphene sheet and the model prohibits the destructive presence of two oxygen atoms on the same hexagon. This simplified model could well describe the functionalization of graphene due to ozone treatment\cite{Lee} and comparison\cite{Leconte} with experimental results\cite{Moser} backs this up.

\section{Numerical Techniques}

The first part of this section presents {\it ab initio} calculations performed to predict the structural properties of epoxy bound oxygen. Likewise, it handles how the TB parameters were extracted from these calculations. The second part of the section sets out the foundations of the Kubo-Greenwood formalism developed in a TB framework.

\subsection{From \textit{\textbf{ab initio}} to tight-binding models}
\label{TBmodel}

The Density Functional Theory (DFT) calculations are conducted using the SIESTA code\cite{Siesta1, Siesta2, Siesta3}, within the local density approximation (LDA) on the exchange-correlation functional in the Ceperley-Alder\cite{ceperley80} form parametrized by Perdew and Zunger\cite{perdew81}. Core electrons are included using Troullier-Martins\cite{troullier91} pseudopotentials. Double $\zeta$ plus polarization orbitals are used to define the basis set. 

\begin{figure}[t]
  \begin{center}\leavevmode
    \includegraphics[width=8cm]{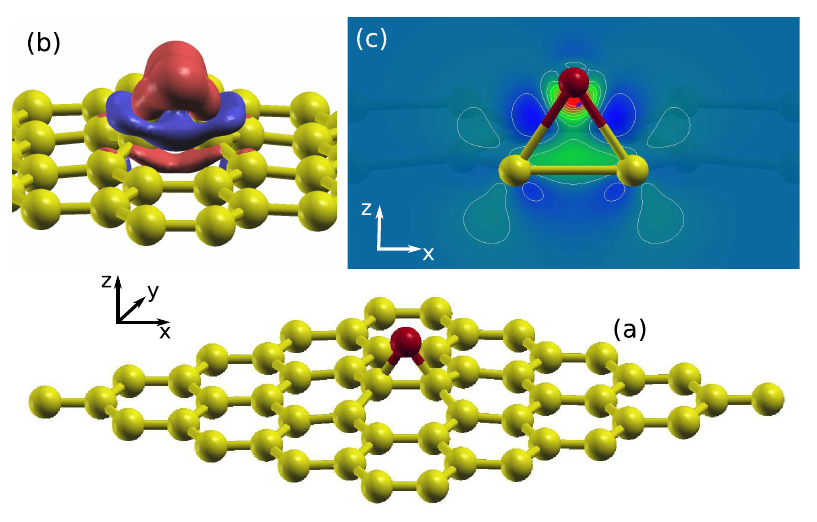}
    \caption{(color online) An oxygen atom in epoxy position on a $5\times5$ cell (a). 3D (b) and 2D (c) charge density difference as defined in the text.
    Charge accumulation in red/green and charge depletion in blue. Isovalues of $0.006$ $|e|/\text{\AA}^3$ and $-0.006$ $|e|/\text{\AA}^3$ for the 3D charge density difference.} 
    \label{fig1}
  \end{center}
\end{figure}

\begin{figure*}[!ht]
  \begin{center}\leavevmode
    \includegraphics[width=16cm]{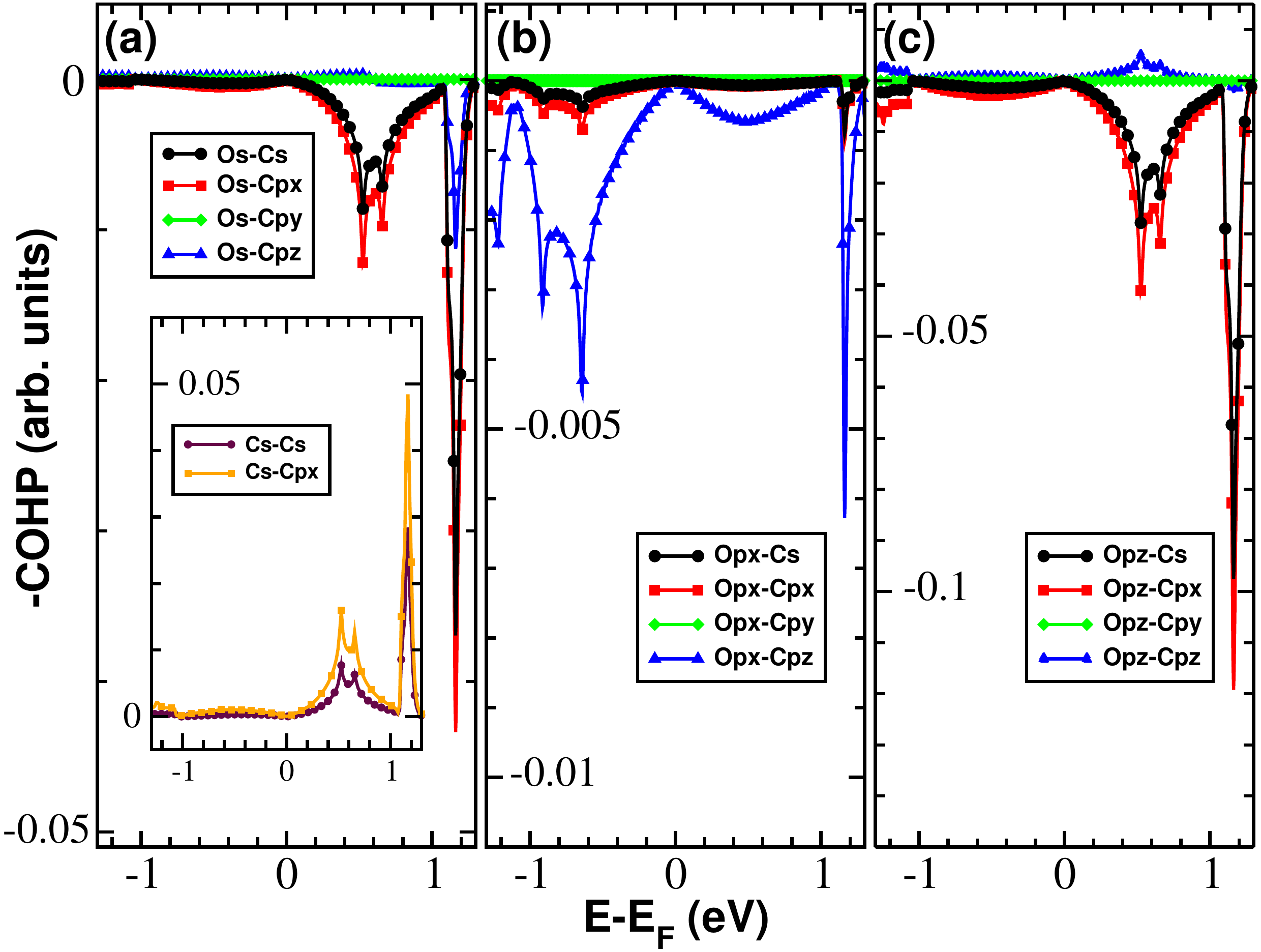}
    \caption{(color online) {\it Ab initio} COHP analysis of the orbitals participating to the bonding of the oxygen atom in epoxy position with its neighboring carbon atoms. Region of
    interest between $-1$ eV and $1$ eV with respect to the Fermi level. Positive and negative values indicate bonding and anti-bonding interactions, respectively.} 
    \label{cohp}
  \end{center}
\end{figure*}

Fig.~\ref{fig1}(a) illustrates the oxygen atom in epoxy position after {\it ab initio} geometry optimization. In order to investigate the chemical bonding of oxygen on graphene, the charge density difference is calculated. Plotting such charge density difference allows for a visual understanding on how the electronic clouds associated with the orbitals are altered by chemical bonding. Fig.~\ref{fig1}(b-c) illustrates this charge density difference $\rho^{\text{diff}}$ defined as follows:
\begin{equation}
  \rho^{\text{diff}} = \rho^{\text{tot}} - \rho^\text{O} - \rho^{\text{graph}}
\end{equation}
where $\rho^{\text{0}}$ and $\rho^{\text{graph}}$ represent the charge densities of freestanding oxygen and graphene respectively. $\rho^{\text{tot}}$ is the charge density of the total system in its bound state. 

In Fig.~\ref{fig1}(b), the charge accumulation in red and the charge depletion in blue indicate a covalent bonding between the carbon and the oxygen atoms. Oxygen is known to exhibit an acceptor behavior (attracts electrons on the $s$ and $p_y$ orbitals). Its $p_y$ orbital does not participate in the bonding as the electronic cloud around the oxygen keeps its conventional $p$-orbital form.

In Fig.~\ref{fig1}(c) the accumulation of electrons is rendered in red and green and the depletion in blue. The typical $\pi$-orbitals of $sp^2$ graphene are broken by the epoxy bonds. Indeed, this bonding causes a charge depletion (blue region) in the $\pi$ electron cloud located above and under the $\sigma$ bond associated with the underlying carbon atoms. The latter $\sigma$ bond thus encounters a slight charge accumulation (green region). Additionally, the electronic charge transfer calculated with a Hirshfeld\cite{Hirshfeld, Fonseca} (Voronoi\cite{Voronoi, Fonseca}) integration adds up to $-0.256$ $|e|$ ($-0.266$ $|e|$) of charge transfer towards the oxygen (red and green region close to the oxygen atom), which indicates that the bonding is also slightly ionic. These two integration techniques yield better results than the frequently used Mulliken integration since they are basis independent\cite{Zwijnenburg}. The oxygen bound in epoxy position is found to perturb graphene, but not in a purely $sp^3$-hybridization way as in the hydrogenation case.

A Crystal Orbital Hamilton Population (COHP) study\cite{COHP,COOP} partitions the band structure energy in terms of orbital pair contributions.
Positive density values represent the bonding states, while negative values describe anti-bonding states when plotting the conventional -COHP. Fig.~\ref{cohp} presents a COHP study on the orbitals participating to the bonding
between the oxygen atom in epoxy position and its neighboring carbon atom. In the following analysis, conclusions are drawn for the region of interest for transport properties $[-1~\text{eV}; 1~\text{eV}]$ only; given bonds may have different bonding or antibonding behaviors away from the Fermi energy. 

\begin{figure}[t]
  \begin{center}\leavevmode
    \includegraphics[width=8cm]{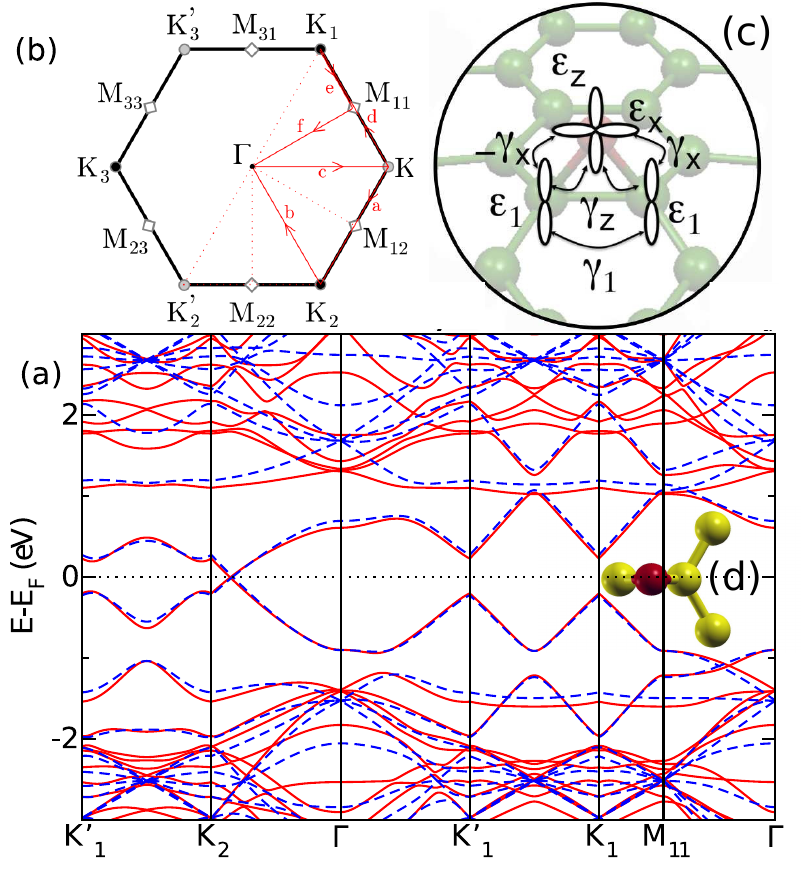}
    \caption{(color online) Electronic band structure of a $5 \times 5$ graphene supercell containing a single epoxy group. The DFT (red-solid) and TB (blue-dashed) band
    structures (a) are described along high-symmetry paths as described in the Brillouin zone (b). Nomenclature of extracted TB parameters is illustrated in (c). One possible orientation of the oxygen
    atom in epoxy position is described herewith. The two other inequivalent orientations are obtained by rotating the oxygen atom in epoxy position by $120\degree$ around the central carbon atom in (d).}
    \label{sym123}
  \end{center}
\end{figure}

One first notes that the contributions of the $s$ (a) and $p_z$ (c) orbital of oxygen bonding with a neighboring carbon are analog. Both these orbitals have a dominant antibonding contribution with the $p_x$ and $s$ orbital of carbon at the right of the Fermi energy. The COHP study between both first neighboring carbon atoms [Fig.~\ref{cohp} (a), inset] shows that these two orbitals of carbon, also responsible for the $\sigma$ bond hybridization, strongly bind at the same energies ($E\sim0.5$ eV and $1.2$ eV), with analog contributions, typical for the $\sigma$-like hybridization between carbon atoms. These four orbitals ($O_{p_z}$, $O_s$, $C_{p_x}$ and $C_s$) thus form a hybridized electronic cloud with bonding contributions between the two carbon atoms and antibonding contributions between oxygen and carbon. This  result is in agreement with the charge density rearrangement observed in Fig.~\ref{fig1}. The electronic charge depletion of the $\pi$ orbitals observed in Fig.~\ref{fig1} can be rationalized with a COHP study for a larger energy window (not shown here). The $\pi$ electrons are partly drawn into the $\sigma$ bond between the carbon atoms. The remaining $p_z$ electrons of carbon [blue lines in (a), (b) and (c)] interact mainly with the $p_x$, $p_z$ and $s$ orbitals of oxygen. Finally, the $p_y$ orbital of oxygen does not interact with carbon in this energy window. 

Consequently, a TB model with first-neighbor interactions including both $p_x$ and hybridized $s$/$p_z$ orbitals of oxygen with the $p_z$ and $s/p_x$ orbitals of carbon should be sufficient to accurately model the effect of epoxy groups on graphene. 
The combined contributions of the three orbitals of carbon binding with oxygen is renormalized to only one orbital in a $\pi$-like model. The matrix elements of the TB Hamiltonian are given by:
\begin{equation}
  \mathcal{H}_{ij} = \sum_{ij} \gamma_{ij} a_i^{\dagger} a_j + \sum_{i} \epsilon_{i} a_i^{\dagger} a_i.
  \label{}
\end{equation}
At first, a $\sqrt{3}\times \sqrt{3}~\text{R}~30\degree$ supercell with one epoxy atom was simulated using DFT to extract a band structure with limited folding of the Brillouin zone (not shown here). The bands near the Fermi energy are nicely fitted using the following TB parameters [nomenclature, see Fig.~\ref{sym123}(c)]: $\epsilon_x = -2.5$ eV, $\epsilon_z = -1.0$ eV, $\epsilon_1 = 1.5$ eV, $\gamma_x = 1.8 \gamma_0$, $\gamma_z = -1.5 \gamma_0$ and $\gamma_1 = 0.0$ eV with $\gamma_0 = -2.6$ eV. 

\begin{figure}[t]
  \begin{center}\leavevmode
    \includegraphics[width=8cm]{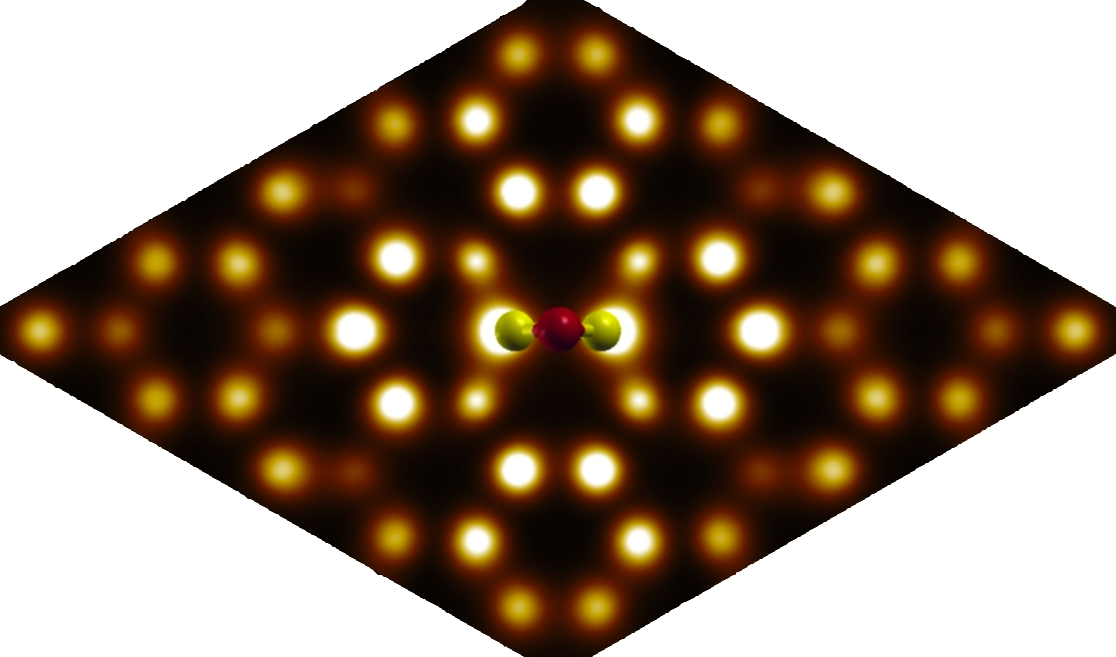}
    \caption{(color online) DFT STM image of empty states, {\it ie.} Local DOS for one oxygen atom in epoxy postion, integrated between $0$ and $0.5$ eV. Both sublattices are affected equally preserving the pseudospin symmetry in graphene.}
    \label{fig4}
  \end{center}

\end{figure}
Finally, to confirm the validity of the model, these TB parameters are used to generate the band structure of a $5 \times 5$ supercell containing a single epoxy oxygen, which is superimposed with its DFT counterpart [see Fig.~\ref{sym123}(a)]. The electronic path chosen to plot the band structure contains all inequivalent high-symmetry segments in the 2D Brillouin zone of a $5 \times 5$ graphene supercell\cite{Lherbier2}[see Fig.~\ref{sym123}(b)]. Note the band crossing at the Fermi energy is shifted away from the $K_2$ point. The localized flat band (visible around the $\Gamma$ point) appearing in the DFT band structure around -2.5 eV is missing in the TB band structure. This originates from a strong interaction between the $p_y$ orbital of oxygen with the $p_y$ orbital of carbon. Both these orbitals are missing in the TB model.
These band structures were calculated for one orientation of the epoxy group on graphene[see Fig.~\ref{sym123}(d)]. For the two other possible orientations of the epoxy oxygen the crossings occur close to $K_1$ or $K_3$ for symmetry reasons. In conclusion, our TB model seems to be sufficient to accurately model random positions and random orientations of impurities, as long as these epoxy oxygens do not interact with each other which is assumed to be satisfied for the range of concentrations of impurities considered here.

Finally, in Fig.~\ref{fig4}, the twofold $D_{2h}$ symmetry in the simulated STM image obtained by integrating the local DOS (LDOS) proves the analogy with a double impurity defect, by comparison with the LDOS in Fig.~2 of Ref.~[\onlinecite{Wehling}]. The LDOS of empty states is spatially integrated between $0$ and $0.5$ eV. A similar pattern (not shown here) is obtained for hole carriers, by integration between and $0$ and $-0.5$ eV. 

\subsection{Kubo Formalism}
Transport properties for large mesoscopic-sized systems can be simulated efficiently using an order-N method based on the Kubo formalism\cite{Kubo1, Kubo2, Kubo3, Kubo4, Kubo5, Kubo6, Kubo7, Kubo8}. Assuming the electronic transport in the system is isotropic for the in plane $x$ and $y$ directions, the 2D diffusion coefficient $D(t)$ is obtained by
\begin{equation}
  D(t) = D_x(t) + D_y(t) = 2D_x(t)
\end{equation}
Within this formalism, the diffusion coefficient $D_x(t)$ in the transport direction $x$ is calculated at each time step using
\begin{equation}
D_x(t) = \frac{\Delta X^2(t)}{t}
\label{eqD}
\end{equation}
where
\begin{equation}
  \Delta X^2 (E,t) = \frac{\text{Tr}\left[\delta(E-\hat{H})\left|\hat{X}(t)-\hat{X}(0)\right|^2\right]}{\text{Tr}\left[\delta(E-\hat{H})\right]}
\label{DeltaX2}
\end{equation}
where $\hat{X}(t)$ is the position operator in Heisenberg representation at time t:
\begin{equation}
  \hat{X}(t) = \hat{U}^{\dagger}(t)\hat{X}(0)\hat{U}(t)
  \label{}
\end{equation}
and where $\hat{U}(t)=e^{-i\hat{H}t/\hbar}$ is the time-evolution operator.

The trace, which is a sum over wavepackets initially localized on each orbital of the system, is replaced by an initial state with a random phase on each orbital of the system. Taking the average of ten initial random phase states already yields very satisfactory results on the smoothness of the curves. This greatly reduces computation time. $U(t)$ can be expanded using Chebyshev polynomials to allow for the mandatory order-N method to achieve reasonable computation time for systems containing
millions of orbitals. Both the numerator and denominator in Eq.~(\ref{DeltaX2}) are calculated using the Lanczos recursion scheme thanks to continued fractions expansions. The termination term is the one usually used for metals, which considers that the oscillation of the recursion coefficients is rapidly damped with the number of recursion steps. We checked that for the recursion step $n = 500$ the damping is sufficient, although a very small remnant oscillation caused by the small energy gap at high is observed. This is quantitatively correct at low energies and qualitatively sufficient at the border of the energy spectrum where the energy gaps occur, by comparison with other more sophisticated termination methods.

From the diffusion coefficient, the mean free path $\ell_e(E)$ and the semi-classical conductivity $\sigma_{\text{sc}}(E)$ can be calculated using respectively:
\begin{equation}
\ell_e(E)=\frac{D^{\text{max}}(E)}{2v(E)}
\label{eqLpm}
\end{equation}
and
\begin{equation}
\sigma_{sc}(E)=\frac{1}{4}e^2\rho(E) D^{\text{max}}(E)
\label{eqKubo}
\end{equation}
where $v(E)$ is the charge carrier velocity at energy $E$, $D^{\text{max}}$ the maximum value of $D(t)$, $e$ the electronic charge, and $\rho(E)$ the DOS at energy $E$. The semi-classical Kubo-Greenwood conductivity $\sigma_\text{sc}$ can be compared to the Drude approximation close to the Dirac point: 
\begin{equation}
\sigma_{\text{D}}(E)=\frac{4e^2}{h}\frac{k(E)\ell_e(E)}{2}
\label{eqDrude}
\end{equation}
where $E = \hbar v_\text{F} k$ with $v_F$ the Fermi velocity close to the Dirac point ($d_{cc}\approx 1.42$ \AA):
\begin{equation}
  v_F \approx \frac{3\gamma_0 d_{cc}}{2\hbar} \approx 1\times 10^6 m s^{-1}
  \label{}
\end{equation}

The limitations of the Drude approximation have recently been discussed and put into context in a Review paper [\onlinecite{Mucciolo}] and the limitations of the Born approximation in the Boltzmann theory of conductivity have been analyzed in Ref.~[\onlinecite{Klos}].

\section{Results}

To begin with, we discuss the effect of various oxygen concentrations on the density of states (DOS) in comparison with pristine graphene. Then, the results obtained for the diffusion coefficient $D(t)$ are analyzed. The other transport quantities are calculated within the Kubo formalism as introduced in the previous paragraph. Particular attention is given to the scaling behavior of the Kubo conductivity. All TB calculations were performed on systems containing 2560000 carbon atoms, which corresponds approximately to systems of $300$ nm by $200$ nm. 

\subsection{DOS and energy shift}

The evolution of the DOS, calculated from $\text{Tr}\left[\delta(E-\hat{H})\right]$, with increasing impurity density is reported on Fig.~\ref{fig5}. 
Although DFT calculations on similar concentrations exhibit a shift of the Fermi energy compared to the pristine case (not shown here), no shifts are applied yet to point out the similarity with other studies\cite{Neto, Wu} covering the effect of ideal impurities ({\it ie.} the influence of the change of respectively the on-site energy and the hopping parameters) on the DOS of pristine graphene.

A shift of the minimum of the DOS is observed with increasing impurity concentrations. Even though formally the rigid band theorem cannot be used for atoms that do not have the same valence ({\it ie.} carbon and oxygen)\cite{Kittel}, the shift is found to be linear with increasing concentrations ($x$) and the second order corrections $O(2)$ can thus be neglected [see Fig.~\ref{shiftedDos} (inset)]:

\begin{figure}[t]
  \begin{center}\leavevmode
    \includegraphics[width=8cm]{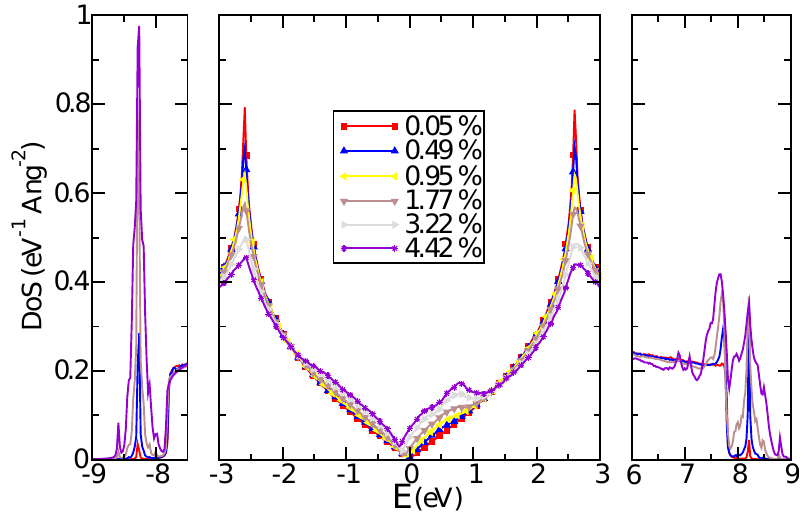}
    \caption{(color online) Main frame: DOS for various impurity densities ranging from $0.05$ to $4.42$ \%. The minimum of DOS shifts slightly with the addend concentration due to
    the changes in hopping parameters and on-site energies in the TB model. Side panels: $\delta$-like peaks corresponding to impurity bands at lower (left) and higher (right) energies.} 
    \label{fig5}
  \end{center}
\end{figure}


\begin{equation}
  \Delta \epsilon_n = xU + \mathcal{O}(2) + ... \ 
\end{equation}
where $U$ is the local potential induced by the epoxy defect. A linear fit of $\Delta \epsilon_n$ versus $x$ implies a value of $U$ equal to $\sim -3.7$ eV. 

\begin{figure}[t]
  \begin{center}\leavevmode
    \includegraphics[width=8cm]{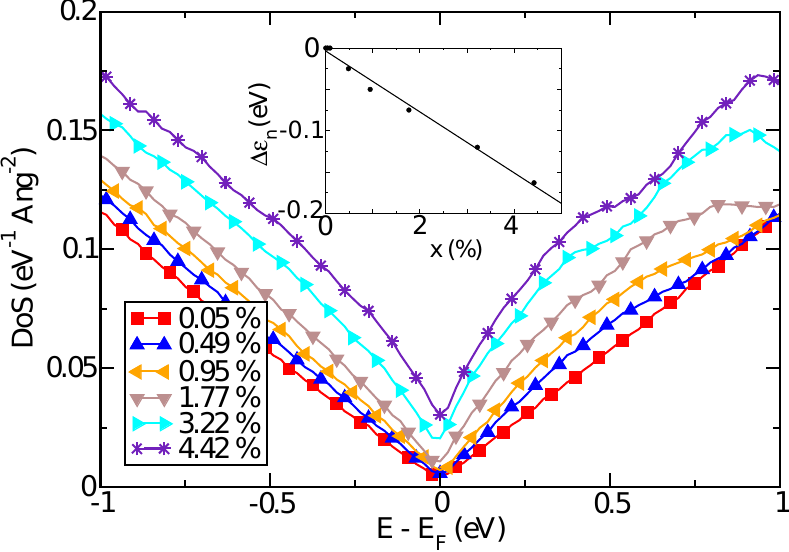}
    \caption{(color online) Main frame: Realignment of the minimum of DOS and charge neutrality point (CNP: position obtained by DOS integration) at $0$ eV. Inset: linear increase of energy shift $\Delta \epsilon_n$ with increasing concentrations of epoxy groups ($x$ in $\%$). Rigid band theorem (see text) implies an impurity induced potential of $\sim -3.7$ eV.} 
    \label{shiftedDos}
  \end{center}
\end{figure}

\begin{figure}[t]
  \begin{center}\leavevmode
    \includegraphics[width=8cm]{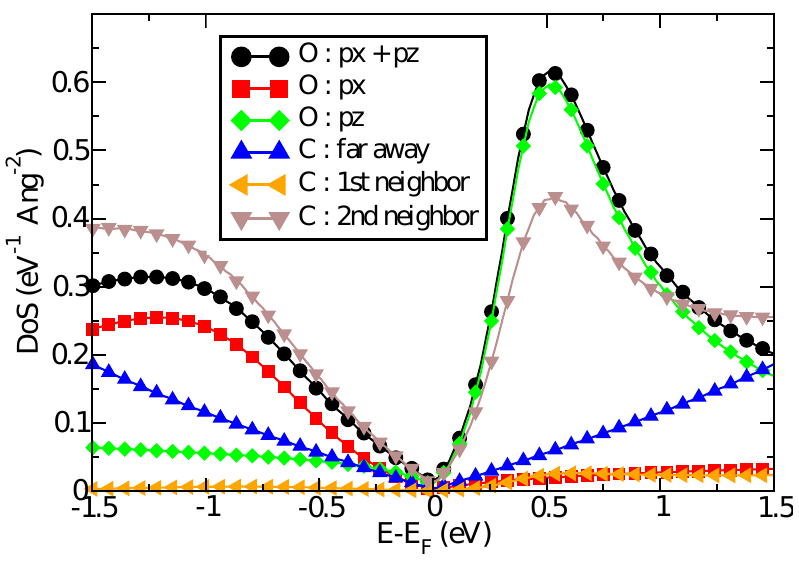}
    \caption{(color online) Projected densities of states. The $p_x$ ($p_z$) orbital of oxygen mainly contributes on the left (right) side of the Fermi energy. Negligible contributions are predicted for first-nearest neighboring carbon atoms. Second carbon nearest neighbors do contribute to the resonant energy bump with oxygen.} 
    \label{PDOS}
  \end{center}
\end{figure}

In addition, the Van Hove Singularities (VHS) are smoothened out and decreased in amplitude with increasing concentrations of epoxide groups, in agreement with previous observations\cite{Kubo1, Neto, Wu}. Also, a small increase of the DOS appears at the minimum of the DOS with increasing impurity concentration.

The bumps in densities of states on the left and on the right of the minimum of DOS (Fig.~\ref{fig5}, middle panel) correspond to the resonant energies between the oxygen atoms in epoxy position and the graphene sheet\cite{Wehling}, while the $\delta$-like peaks in the side panels at high energies correspond to flat impurity bands. As discussed in Section~\ref{TBmodel}, the localized state close to the left VHS caused by the strong interaction between both $p_y$ orbitals of oxygen and carbon is missing in this simplified TB model.

Finally, oxygen in epoxy position triggers a shift of the Fermi energy which compensates the shift of the minimum of DOS discussed above (see Fig.~\ref{shiftedDos}). This Fermi energy is obtained by integrating the TB DOS and counting the number of electrons present in the system. 
In the next Sections, transport calculations will implicitly include this realignment of minimum of DOS with the Fermi energy, thus locating the charge neutrality point (CNP) at $E=0$ eV.

In Fig.~\ref{PDOS}, a Projected Densities of States (PDOS) evidence that the bump on the right side of the Fermi
level originates from the $p_z$ orbital and the one on the left from the $p_x$
orbital of oxygen, in agreement with the previous COHP DFT study. Fig.~\ref{PDOS}
also indicates that the first-nearest neighboring carbon atoms do not contribute to the
total DOS in contrast to the second nearest neighboring carbon atoms. The PDOS suggest the
oxygen atom attracts most of the electronic density and thus weakens the density on the first-neighboring carbon. Such analysis agrees with existing literature\cite{Wehling2, Mkhoyan}.

\subsection{Diffusion coefficient and transport regimes}

\begin{figure}[t]
  \begin{center}\leavevmode
    \includegraphics[width=8cm]{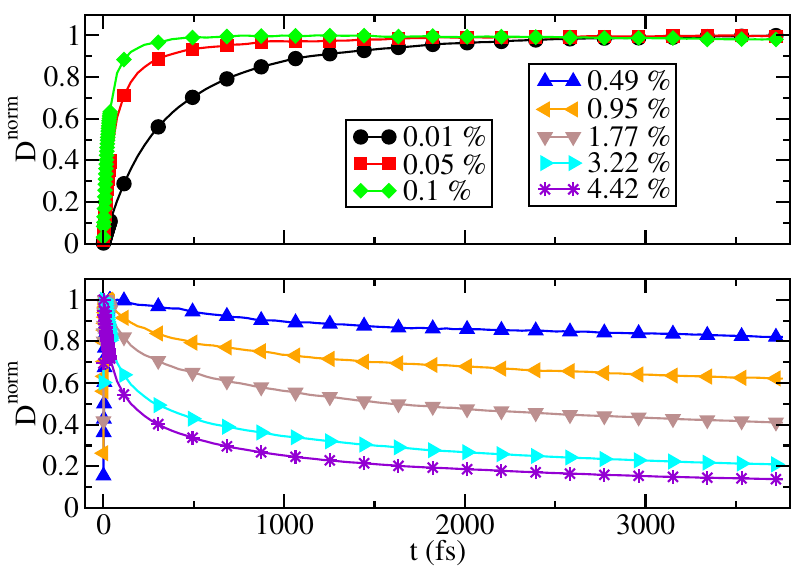}
    \caption{(color online) Time evolution of diffusion coefficient (normalized) at $E = 0.5$ eV for different impurity concentrations.}
    \label{DiffusionConcentration}
  \end{center}
\end{figure}

The diffusion coefficient $D(t)$ inherently contains all the information needed to calculate transport properties [see Eqs.~(\ref{eqLpm}) and (\ref{eqKubo})]. Its time evolution or dynamics also clarifies the dominant transport regime at the considered time scale ({\it ie.} ballistic, diffusive or localized). Fig.~\ref{DiffusionConcentration} illustrates the typical behaviors of the normalized diffusion coefficients
\begin{equation}
  D^\text{norm}(t)=\frac{D(t)}{D^\text{max}}
  \label{}
\end{equation}
for an energy $E = 0.5$ eV. As expected, the conduction in pure graphene is simply ballistic. The smallest simulated impurity concentration ({\it ie.} $0.01$ \%) does not
reach its maximum for the diffusion coefficient for the total elapsed time considered here (approx. $3800$ fs). The slope of $D(t)$ gives access to a $v^2(E)$ value still in close agreement with the theoretical
(analytical) value $v_F^2$ for energies $E$ close to $E_F$. For instance, at $0$ eV and for $0.01$ \%, $v(E) = 2.11 \text{\AA} \gamma_0 / \hbar$, while $v_F = 2.13 \text{\AA} \gamma_0 / \hbar$. For intermediate densities ($0.05$
and $0.1$ \%) $D^\text{norm}(t)$ evolve from the ballistic regime to the diffusive regime for the simulated times. At higher concentrations ($0.49$ \%) $D^\text{norm}(t)$ reach a diffusive regime sharply, followed by a clear decrease with time, signature of quantum interferences leading the charge carrier localization. 

\begin{figure}[t]
  \begin{center}\leavevmode
    \includegraphics[width=8cm]{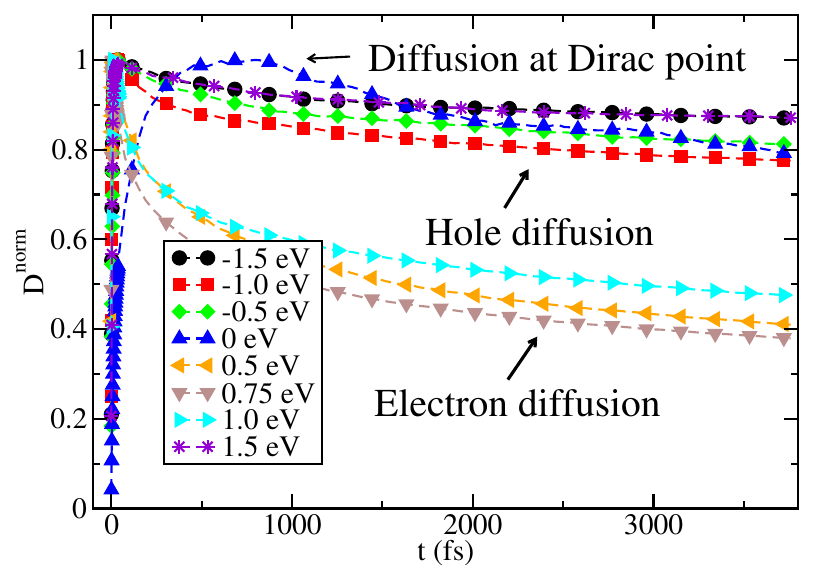}
    \caption{(color online) Normalized diffusion coefficient versus time for $x = 1.77$ \% at different energies.} 
    \label{DiffusionEnergy}
  \end{center}
\end{figure}

Fig.~\ref{DiffusionEnergy} presents diffusion coefficients $D^\text{norm}(t)$ of $1.77$ \% of impurities at different energies. Localization effects follow an asymmetric behavior between electrons and holes. The predominant resonant peak at the right of the charge neutrality point (CNP) causes much stronger localization effects than the weaker and more smeared out peak at the left of the CNP (in analogy with the PDOS in Fig.~\ref{PDOS}). Such asymmetry is of crucial importance to synthesize possible electronic devices made of functionalized graphene requiring an efficient switch between a conducting and an insulating behavior\cite{Biel1, Biel2}. Localization effects are more significant around $0.75$ eV. One finally notes the peculiar behavior of the diffusion coefficient exactly at the CNP. Numerically, this point is more problematic to simulate since the density of charge carriers may be very scarce. Nevertheless, our results suggest that the saturation limit of the Diffusion coefficient is reached for longer simulation time and then turns into a moderate decrease characteristic of quantum effects.

This discussion on the diffusion coefficient points out two different regimes which are approached separately in the remaining Sections. Firstly, the semi-classical quantities, which neglect quantum effects, are discussed in Section \ref{semi}. Secondly, the localization regime is analyzed in Section \ref{quantum}. 

\subsection{Semi-classical regime}
\label{semi}
\subsubsection{Analysis of elastic mean free paths}

\begin{figure}[t]
  \begin{center}\leavevmode
    \includegraphics[width=8cm]{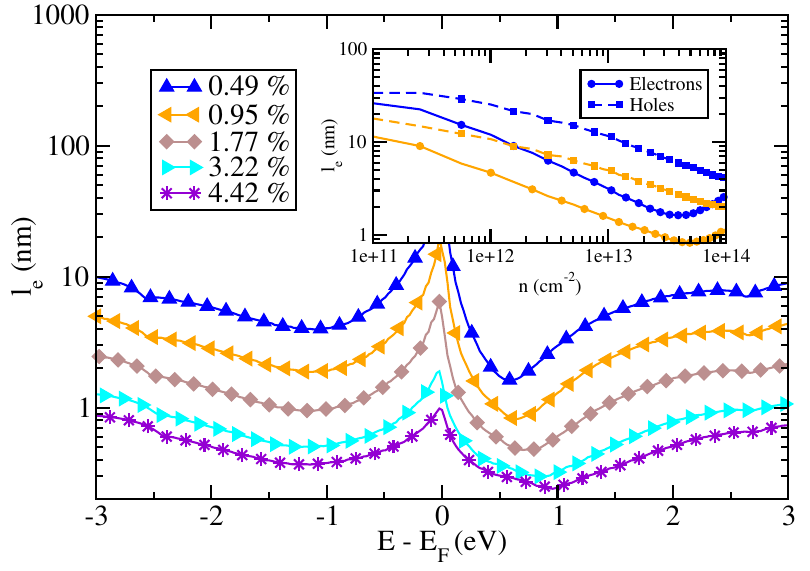}
    \caption{(color online) Energy dependent mean free paths for different impurity concentrations. Inset: mean free path versus hole (dashed/squares) and electron (solid/circles) carrier density for epoxy concentrations of $0.49$ \% and $0.95$ \%.} 
    \label{lpm}
  \end{center}
\end{figure}

\begin{figure}[t]
  \begin{center}\leavevmode
    \includegraphics[width=8cm]{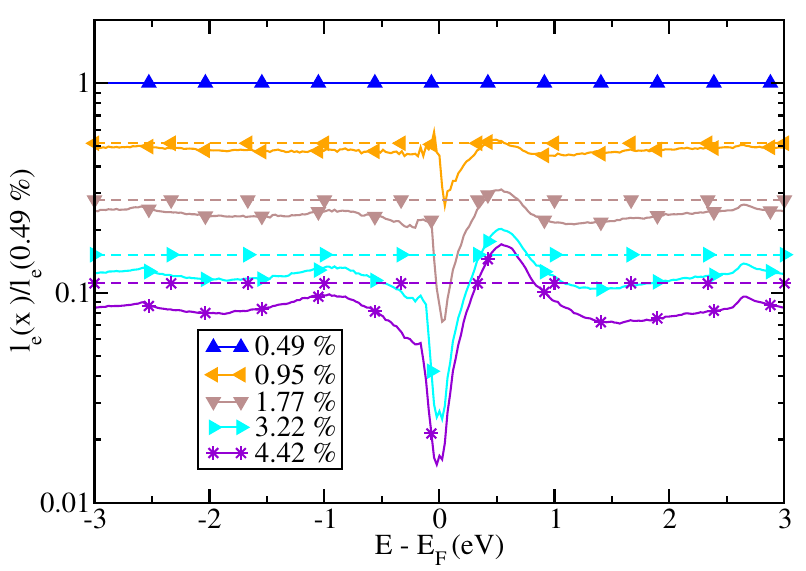}
    \caption{(color online) Ratio of mean free paths at two selected impurity densities. $\ell_e(x_1)/\ell_e(x_2)$ (solid lines) and $x_2/x_1$ (dashed lines) with $x_2 = 0.49$ \%.}
    \label{fermi}
  \end{center}
\end{figure}

Using Eq.~(\ref{eqLpm}), the mean free path $\ell_e$ is calculated using the diffusion coefficient and plotted in Fig.~\ref{lpm} for energies between the two VHS at $-2.6$ eV and $2.6$
eV. For the considered elapsed times, the diffusive regime is not reached at every energy for the smallest impurity concentrations ({\it ie.} $n_i < 0.1$ \%, see Fig.~\ref{DiffusionConcentration}, upper
panel). Therefore, the mean free path can not be estimated for these small concentrations. For concentrations larger than $0.5$ \%, the diffusive regime is reached within the entire energy window. The dip in the mean free path at the right of the CNP (and in a lesser extent, at the left of the CNP), indicating larger scattering effects, shifts away from the CNP and becomes smoother with increasing concentration of oxygen atoms in epoxy position. Such dips correspond to the resonance peaks found in the DOS which are induced by the oxygen. The inset of Fig.~\ref{lpm} confirms the predicted asymmetry, affecting the electrons more strongly. The largest concentrations scatter more uniformly across the entire energy window. In addition, the evolution of the mean free path with impurity concentration follows a simple scaling law as expected from a Fermi golden rule (Fig.~\ref{fermi}):
\begin{equation}
  \frac{\ell_e(x_1)}{\ell_e(x_2)}=\frac{x_2}{x_1}
  \label{FermiRule}
\end{equation}

\begin{figure}[t]
  \begin{center}\leavevmode
    \includegraphics[width=8cm]{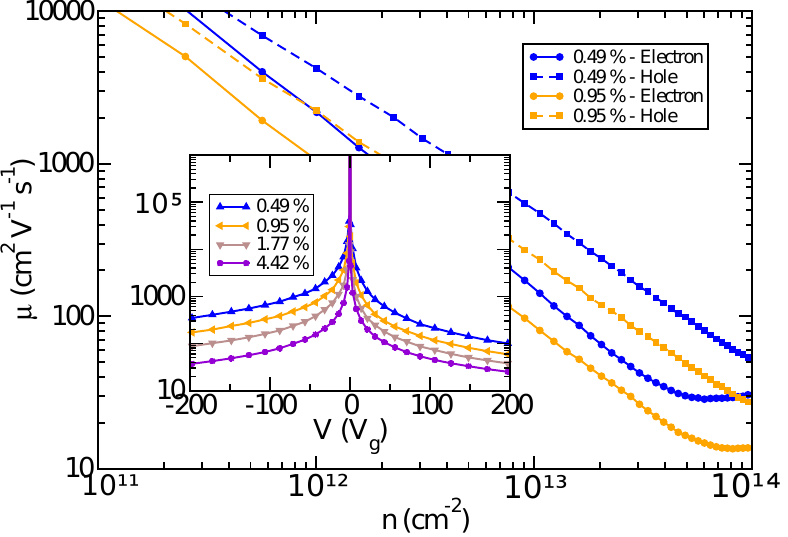}
    \caption{(color online) Main frame: electron and hole mobility for usual experimental carrier densities. More severe scattering effects for negative charge carriers cause an asymmetry in mobility
    compared to the hole mobility. Inset: same mobilities with respect to the gate voltage $V_g$.}
    \label{mobility}
  \end{center}
\end{figure}

\subsubsection{Mobility}

In Fig.~\ref{mobility} (mainframe) the mobility of the charge carriers is estimated theoretically using: 
\begin{equation}
  \mu(E) = \frac{\sigma_\text{sc}(E)}{ne}.
  \label{eqMob}
\end{equation}

Scattering effects are affecting the electron mobility more strongly than the hole mobility. This asymmetry is reduced for the largest impurity concentrations [see Fig.~\ref{mobility} (inset)]. 
 
Experimentalists usually consider the absolute value of the mobility as a key quantity to characterize samples and corresponding inherent disorder. Temperature breaks the phase coherence of electrons along the scattering path and generally reduces quantum interference effects. Accordingly, the use of the semi-classical conductivity in evaluation of $\mu(E)$ (Eq.~\ref{eqMob}) is a reasonable approximation to analyze the experimental data. On the same basis, computed semi-classical conductivities are expected to be more valuable for comparison with conductivities measured experimentally at room temperature. One may argue that in such a non-zero temperature environment, electron-phonon coupling may play also a significant role. However, inelastic scattering lengths due to electron-phonon coupling are extremely long in graphene and may thus be disregarded too, at least as a first approximation.

\subsubsection{Numerical Kubo conductivity and Drude approximation}

\begin{figure}[t]
  \begin{center}\leavevmode
    \includegraphics[width=8cm]{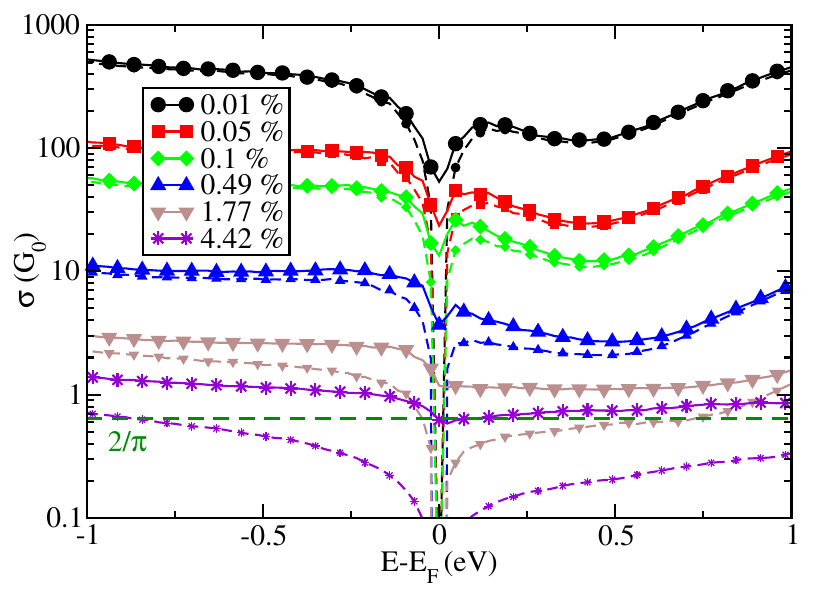}
    \caption{(color online) Comparison between the numerical Kubo conductivity $\sigma_\text{sc}$ (solid lines) and the Drude approximation $\sigma_\text{D}$ (dashed lines) for different impurity concentrations. Minimum theoretical value $2/\pi G_0$ is plotted in horizontal dashed green line.} 
    \label{DrudeVSKubo}
  \end{center}
\end{figure}

Fig.~\ref{DrudeVSKubo} compares the semi-classical value of the Kubo-Greenwood conductivity $\sigma_\text{sc}$ (solid lines) with the Drude conductivity $\sigma_\text{D}$ (dotted lines), extracted from Eq.~(\ref{eqKubo}) and (\ref{eqDrude}) respectively. The Drude approximation seems to be valid only for small
impurity concentrations in the energy window $[-1~\text{eV}; 1~\text{eV}]$. For larger densities, the conductivities are underestimated by a factor of two for energies close to the CNP. Moreover, $\sigma_\text{D}$ is ill-defined at the CNP [see Eq.~(\ref{eqDrude})]. Indeed, at the CNP, $k = 0$ and consequently, $\sigma_\text{D}$ vanishes. However, theoretical work\cite{WALTHEO2,Ando,Kubo2} reports extensively on a minimum value of the semi-classical conductivity equal to $2/\pi G_0=4e^2/\pi h$ (when neglecting quantum interferences). To avoid the singularity at $0$ eV, the mean free path could be calculated analytically, including its $k$ dependence explicitly. As illustrated by the calculations for larger impurity concentrations, the variations of the DOS with disorder have to be included, as it is presently the case in the Kubo formula [Eq.~(\ref{eqKubo})]. These variations account for single and multiple scattering events mentioned in Ref.~[\onlinecite{Santos}].  Up to some numerical discrepancy, the semi-classical minimum value of conductivity is confirmed. 

Additionally, an asymmetry exists between the electron and hole conductivities.
This effect is weakened but not cancelled with increasing impurity
concentrations. This is similar to the asymmetry already observed for both charge carriers in their respective mean free paths and mobilities. Albeit physically different, the stronger
quantum localization effects on the electron side are directly linked to the
stronger classical scattering effects on the same energy window (see Fig.~\ref{DiffusionEnergy}). This will be emphasized in Section \ref{quantum} by studying the evolution of the conductivity with time deep into the diffusive regime.

\subsubsection{Short-range \textit{\textbf{vs}} long-range scattering}
The nature of scattering range induced by oxygen atoms placed in epoxy position can be discussed using the scaling properties of semi-classical quantities. Both the scattering time and the conductivity are here briefly outlined\cite{Placais}.

The elastic scattering time is defined by 
\begin{equation}
  \tau(E) = \frac{\ell_e(E)}{v(E)}.
  \label{}
\end{equation}
Using the dispersion relation $E=\hbar v_F k$, 
$\tau(E)$ can be estimated in terms of the Fermi wave vector $k$. According to Nomura {\it et al.} [\onlinecite{Nomura}] who also used the Kubo-Greenwood approach, following conclusions apply. Long range (lr) and short range (sr) disorder scattering times scale respectively as follows:

\begin{equation}
  \tau_{\text{lr}} \propto k  \text{     and     }  \tau_{\text{sr}} \propto \frac{1}{k}
  \label{}
\end{equation}

\begin{figure}[t]
  \begin{center}
    \includegraphics[width=8cm]{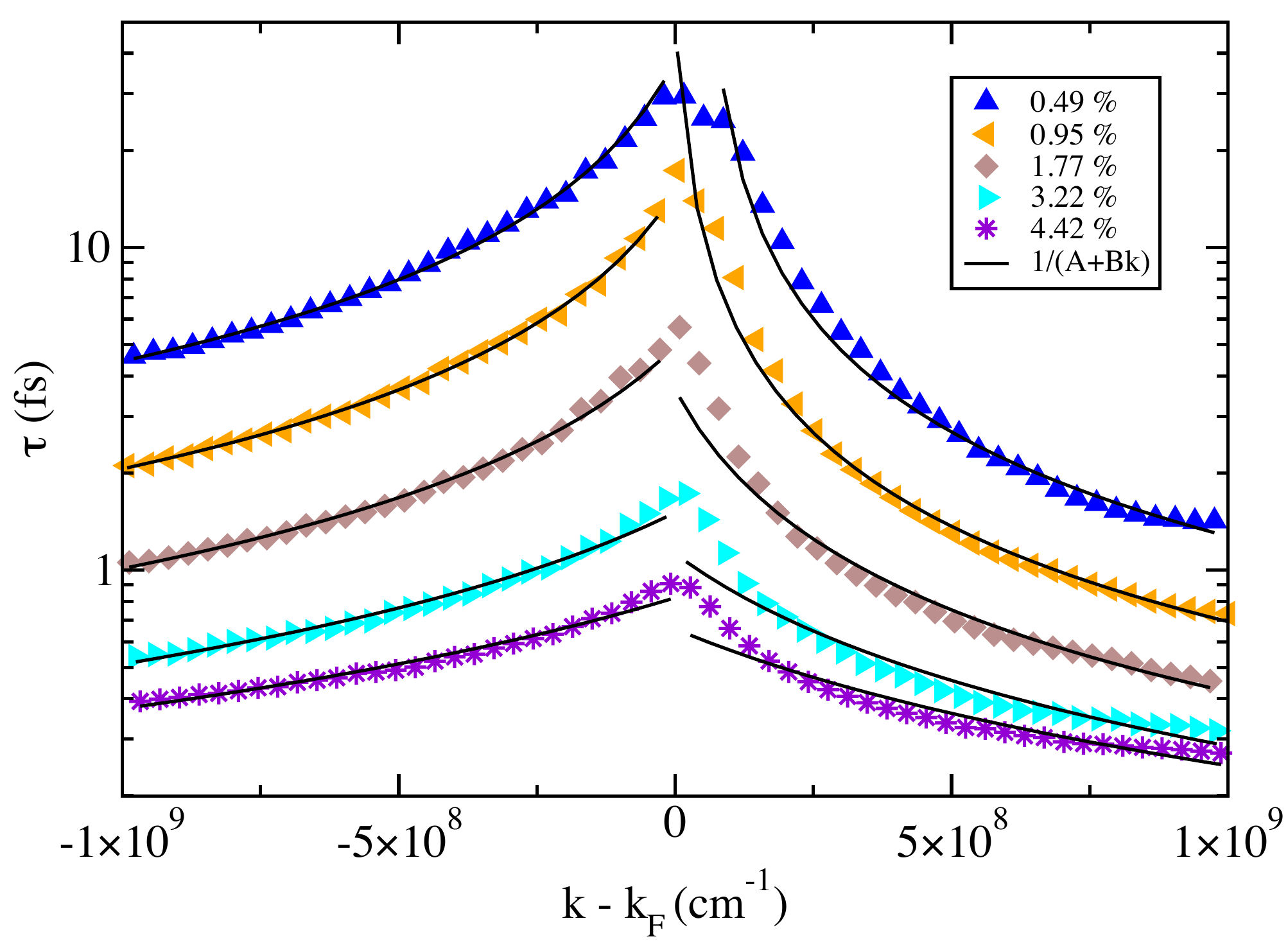}
  \end{center}
  \caption{(color online). Scattering times in function of Fermi wave vector $k$ for different oxygen densities. Numerical fits for each data set based on a regression algorithm with fitting parameters A and B (black) to account for the generalized inverse function. Hole and electron scattering times were fitted separately.}
  \label{scatDens}
\end{figure}

Following such criterion, our data with corresponding numerical fits (Fig.~\ref{scatDens}) clearly indicate a short-range scattering behavior of oxygen in epoxy position.
Such short-range scattering time should diverge at the CNP. This does not happen within the Kubo formalism since the DOS remains finite close to the CNP, in contrast with the prediction of the Drude approximation.

A similar analysis based on the scaling of conductivity is not straightforward.
According to Ref.~[\onlinecite{Nomura}], long range conductivity should scale linearly with $n$, while short range conductivity should display a non-linear behavior, approaching the constant Boltzmann (or Drude) conductivity $\sigma_\text{D}$ for $|E_F| \gg \hbar/\tau$. 
In Fig.~\ref{DrudeVSKubo}, the behavior close to the Dirac point behaves differently depending on the impurity concentration. For smaller impurity concentrations, the conductivity tends to decrease for energies corresponding to reasonable carrier concentrations (up to $10^{13}\text{cm}^{-2}$), while it increases slightly for larger impurity concentrations. The Drude conductivity never reaches a constant plateau for the whole of the energy $E$ or carrier $n$ range. The short-comings of this conductivity have already been pointed out. 

We note that this subject is still debated, as other models\cite{Stauber, Klos, Katsnelson, WALTHEO2} predict a dominant linear dependency (with logarithmic corrections) for both long range and short range conductivity.

Additionally, it has been observed experimentally\cite{Bolotin} that a quasi-ballistic regime exhibits a non-linear behavior with $n$, while a more disordered system scales linearly with $n$. Comparison with experiment becomes particularly tricky as most of previously mentioned analytical models disregard important multiple scattering effects on the computed conductivity. 

Another important remark is that most theoretical predictions have been derived assuming restrictions on disorder models which are partly invalidated in the present study. Indeed, the epoxy defects have been derived from accurate first-principle calculations, and the resulting TB model brings more realism and generality when compared to simplified academic models.
As a matter of fact, the DOS (Fig.~\ref{shiftedDos}) evidences resonant energy bumps, driven by randomly distributed oxygen, which could cause squared logarithmic corrections\cite{WALTHEO2}. Our data cannot really be accurately fitted to obtain the different corrections to the scaling in this context.

This dominant short-range disorder is at the root of the quantum effects presented in the remaining Sections.

\subsection{Evolution of the Kubo conductivity with time scale (or length)}
\label{quantum}

\begin{figure}[t]
  \begin{center}\leavevmode
    \includegraphics[width=8cm]{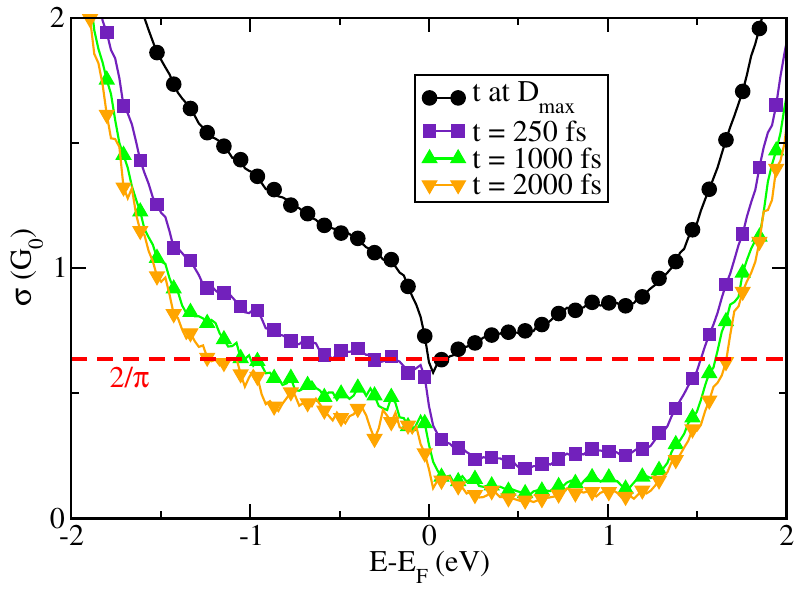}
    \caption{(color online) Kubo conductivities at different elapsed times of wave packet evolution for $4.42$ \% of impurities.}
    \label{kuboWithTime}
  \end{center}
\end{figure}
The semi-classical expression of the Kubo-Greenwood conductivity [Eq.~(7)] restricts the transport to the diffusive regime, {\it ie.} when suppressing quantum interferences.
To follow the time evolution spreading of the quantum wave packets, the expression of $D_x(t)$ in Eq.~(3) should be replaced as follows:

\begin{equation}
  D_x(t) = \frac{\partial (\Delta X^2(t))}{\partial t}
  \label{deriveD}
\end{equation}
When replacing $D^\text{max}$ by the new expression of $D(t)$ [Eq. (\ref{eqD})] in Eq.~(\ref{eqKubo}), the Kubo conductivities are
obtained at different time scales as depicted in Fig.~\ref{kuboWithTime} for a $4.42$ \% impurity density. The time-evolution of the conductivity is found not to be uniform over the energy spectrum, indicating there might be different transport regimes depending on the charge carrier energies and impurity concentrations for a given length scale.

\begin{figure}[t]
  \begin{center}
    \includegraphics[width=8cm]{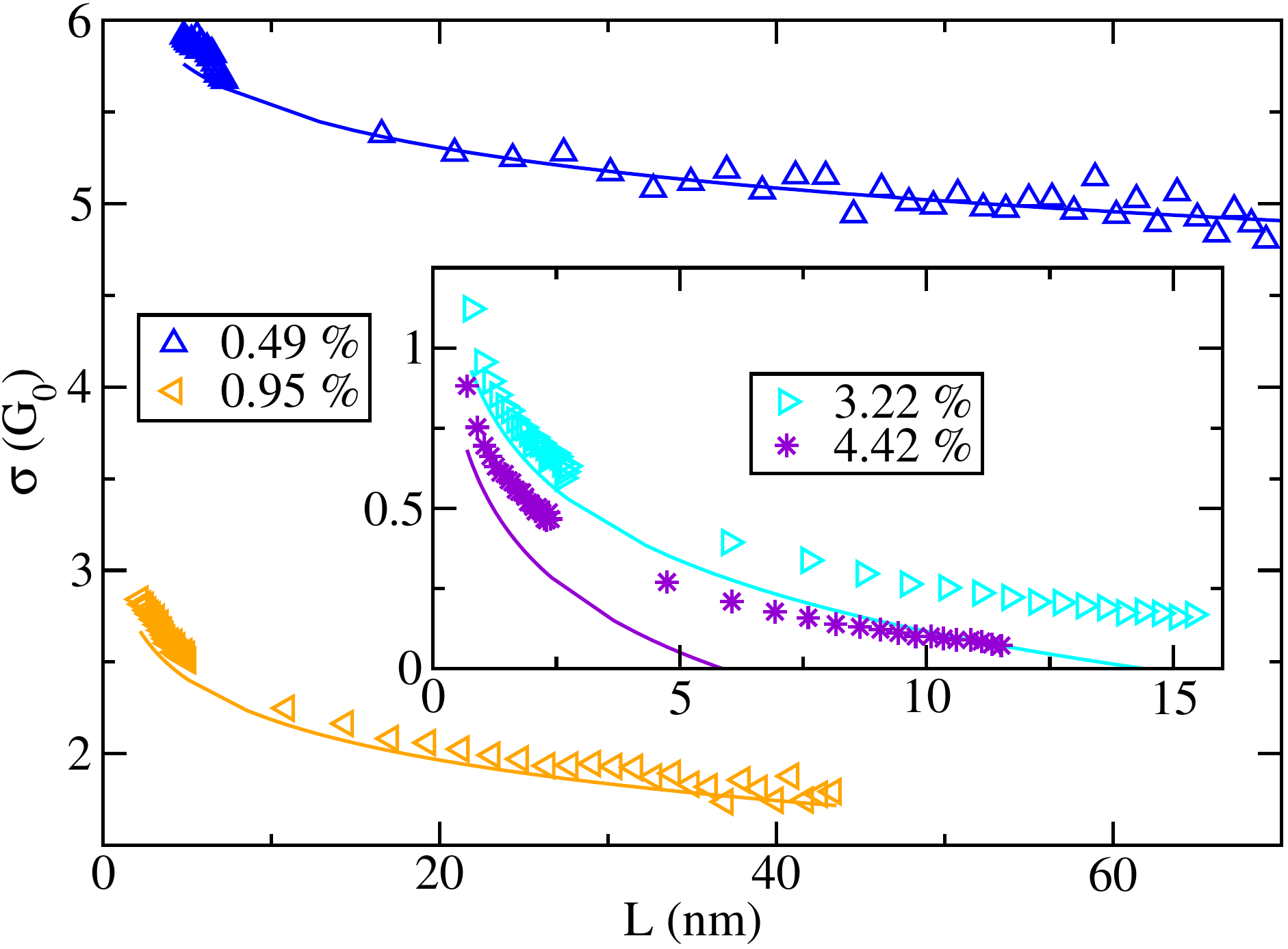}
  \end{center}
  \caption{(color online) Kubo conductivities for $L>L^{\text{max}}$, which include weak localization corrections to the semi-classical conductivity, for different impurity densities at energy $0.8$ eV. The numerically estimated conductivity $\sigma(L)$ (obtained using Eqs.~\ref{deriveD} and \ref{diffL}, symbols) contains numerical jiggling caused by the very sensitive derivation in Eq.~\ref{deriveD}.
  The conductivity obtained using Eq.~(\ref{logL}) is plotted in solid lines. Only one out of five points are plotted in the inset for clarity reasons.}
  \label{fitSigmaL}
\end{figure}

According to the scaling theory of localization\cite{Ramakrishnan},
there are two possible behaviors for the conductivity corresponding to the weak
and strong localization regimes which read as follows:

\begin{equation}
  \sigma(L) = \sigma  \big |_{D^{\text{max}}} - \frac{e^2}{\hbar \pi^2}
  \ln\left(\frac{L(t)}{\sqrt{2} \ell_e}\right)
  \label{logL}
\end{equation}
\begin{equation}
  \sigma(L)\sim \exp\left(-\frac{L(t)}{\xi}\right)
  \label{expL}
\end{equation}
where the localization length $\xi$ gives an estimation of the distance covered by a charge carrier before it is totally trapped due to this multiple scattering effects. The diffusion length is defined as
\begin{equation}
  L(t) = 2 \sqrt{2 \Delta X^2(t)}.
  \label{diffL}
\end{equation} 
This definition of L is reasonable when saturation of the diffusion coefficient has been reached. The extra factor $\sqrt{2}$ in Eq.~\ref{logL} compared to the correction obtained by Lee {\it et al.}~[\onlinecite{Ramakrishnan}] comes from a different definition of $D(t)$.
Both numerical estimation of $\sigma(L)$ (symbols) and analytical $\left[ \sigma  \big |_{D^{\text{max}}} - e^2/\hbar \pi^2  \ln\left(\frac{L(t)}{\sqrt{2} \ell_e}\right)\right]$ from Eq.~(\ref{logL}) (solid lines) are plotted in Fig.~\ref{fitSigmaL} for $L>L^{\text{max}}$. The numerical part contains small jiggling caused by the very sensitive derivation in Eq.~(\ref{deriveD}).

\begin{figure}[t]
  \begin{center}\leavevmode
    \includegraphics[width=8cm]{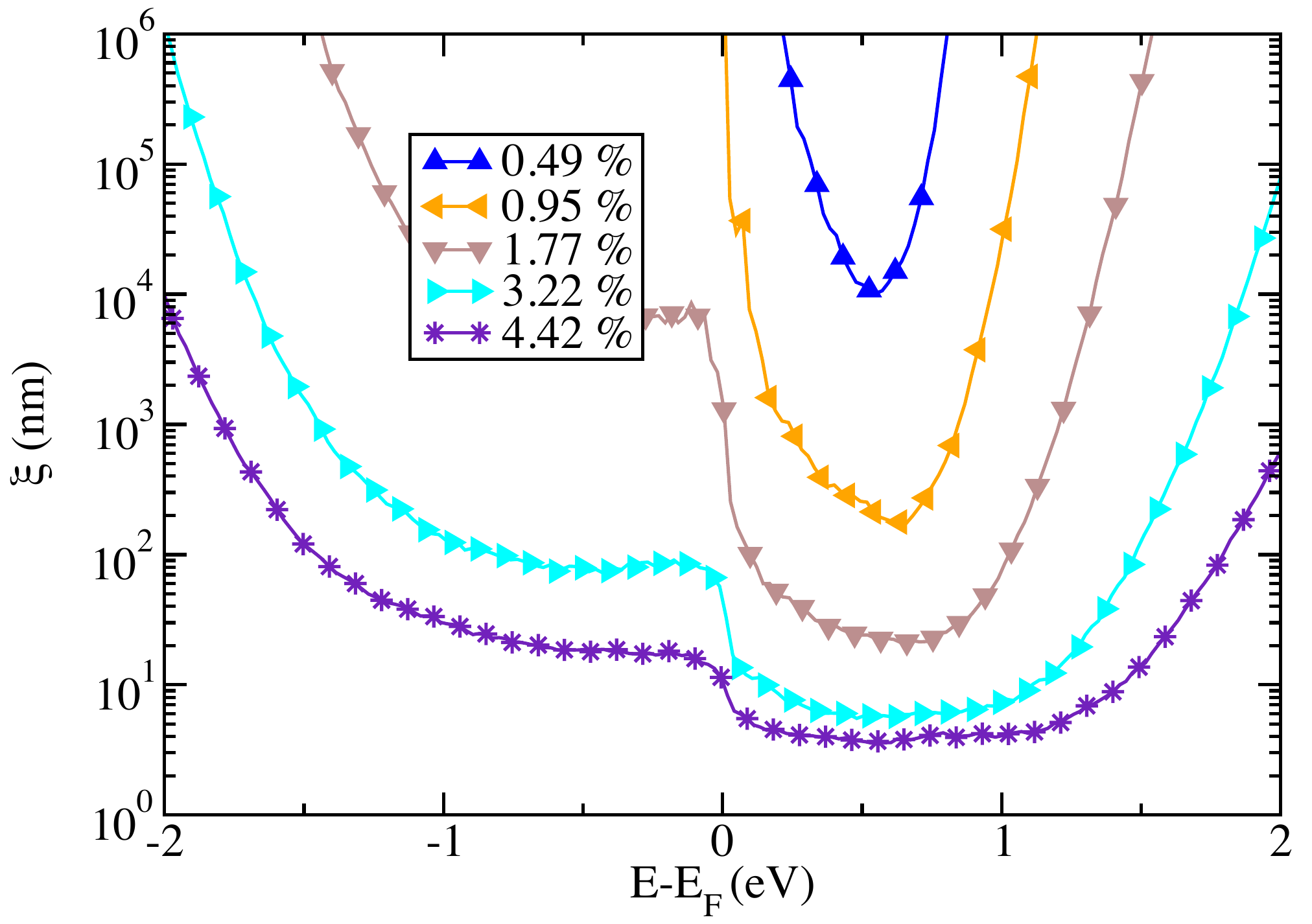}
    \caption{(color online) Localization lengths estimated using Eq.~(\ref{xi}) for different impurity concentrations.}
    \label{xi}
  \end{center}
\end{figure}

The corrections to the semi-classical conductivity in the low impurity limit (mainframe) follows the logarithmic behavior, from which an estimation of $\xi$ can be deduced. $\xi$ corresponds to the length where the cooperon corrections equal the semi-classical conductivity\cite{Ramakrishnan}. Starting from Eq.~(\ref{logL}), the localization length is thus estimated with 
\begin{equation}
  \xi(E)=\sqrt{2} \ell_e \exp\left(\frac{\pi\sigma_\text{sc}}{G_0}\right)
  \label{eqXi}
\end{equation}
with the computed $\ell_e$ and $\sigma_\text{sc}$ and corresponds to the 2D generalization of the Thouless relationship\cite{Thouless, Beenakker}. These localization lengths are plotted in Fig.~\ref{xi}\cite{comment1}.

For larger concentrations, the cooperon corrections to the semi-classical conductivity seem to saturate and depart from the perfect logarithmic behavior (Fig.~\ref{fitSigmaL}, inset). The corrections obtained numerically become smaller than what is predicted due to a transition to the strongly localized regime following an evanescent exponential behavior. This can be rationalized invoking the Ioffe and Regel criterion\cite{Ioffe} which states that the appearance of strong localization becomes significant for impurity concentrations satisfying $k_F \ell_e = 1$. Such a criterion implies a mean free path of the order of the interatomic distance, which is actually the case for $4.42$ \% of impurities where the mean free path is estimated to be below $3 \text{\AA}$ for the energy window from $0.5$ eV to $1.0$ eV. 

\begin{figure}[t]
  \begin{center}
    \includegraphics[width=8cm]{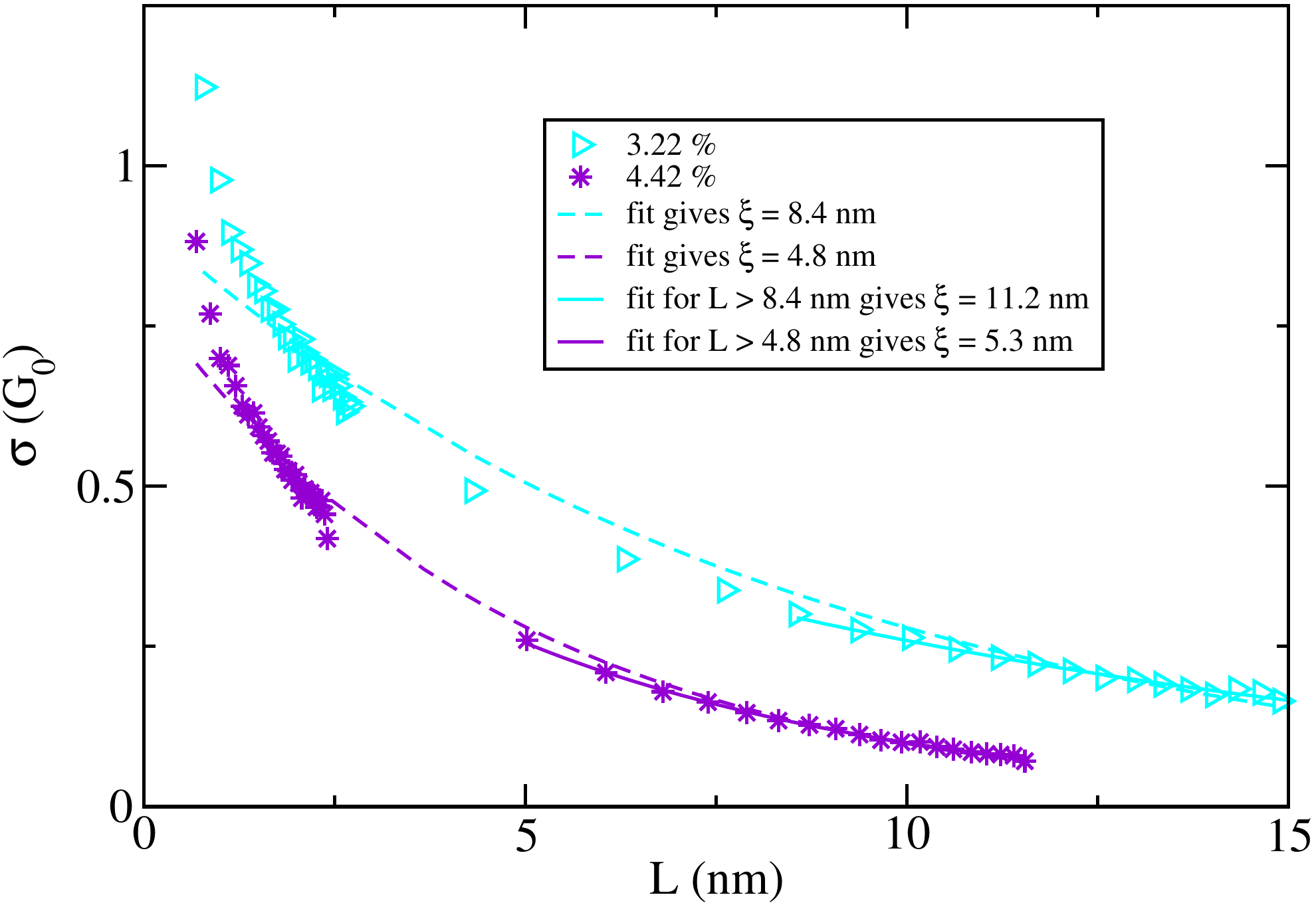}
  \end{center}
  \caption{Exponential fits [Eq.~(\ref{expL})] to estimate localization lengths $\xi$ in the strongly localized regime at energy $0.8$ eV. A first fit (dashed lines) for the whole of the available data allows us to estimate a length L for which wavepackets are localized. A second fit (solid lines) for values larger than L gives us new estimates for $\xi$.}
  \label{fitExp}
\end{figure}

In Fig.~\ref{fitExp}, by fitting the exponential behavior of Eq.~(\ref{expL}), values for $\xi$  equal to $8.4$ and $4.8$ nm are estimated for $3.22$ \% and $4.42$ \% of impurities respectively at an energy of $0.8$ eV (dashed lines). Refitting $\sigma(L)$ for the region at the right of these values (solid lines), we obtain convincing exponential decays and more accurate estimates for $\xi$ equal to $11.2$ and $5.3$ nm, respectively. Both these estimates and the ones obtained by Eq.~(\ref{eqXi}) in Fig.~\ref{xi} are of the same order of magnitude, thus validating our results. Experimentalists however often use the Drude approximation $\sigma_D$ instead of the correct semi-classical conductivity $\sigma_\text{sc}$ in Eq.~(\ref{eqXi}). The inaccuracy of the Drude approximation for largest impurity concentrations causes the localization length to be underestimated by an order of magnitude.

\section{Conclusions}
In this paper, the quantum transport properties of chemically damaged two-dimensional graphene based structures have been investigated. Using the Kubo-Greenwood transport framework, and by means of an efficient order N numerical implementation, mesoscopic transport features in disordered graphene have been explored in details, with impurities (adsorbed oxygen-driven epoxide defects) described by local tight-binding parameters, deduced from first-principles calculations.

In addition to the numerical calculation of the energy-dependent elastic mean free path driven by a given epoxide density, quantum localization effects have been analyzed from the weak to the strong (Anderson) localization regimes. By applying the conventional scaling theory of localization, the 2D-localization lengths have been evaluated from the scaling behavior of the Kubo conductivity, and contrasted to the prediction deduced from the cooperon correction to the conductivity (which relates $\xi$ to the elastic mean free path and semi-classical conductivity). A very reasonable agreement has been obtained, pinpointing further towards a strong energy-dependence of all transport length scales.

By combining the ab-initio approach for the description of the defects structure and local energetics with an efficient and exact quantum transport methodology implemented on tight-binding models, our general theoretical framework provides a solid foundation and tool to understand the origin of complex transport phenomena in strongly disordered and chemically complex graphene-based nanostructures. The extension of our study to any other kinds of defects (topological, chemical, etc) and other types of two-dimensional structures is straightforward.

\acknowledgements
J.-C.C., N.L., A.L. and F.V. acknowledge financial support from the F.R.S.-FNRS of Belgium. This work is directly connected to the Belgian Program on Interuniversity Attraction Poles (PAI6) on 'Quantum Effects in Clusters and Nanowires', to the ARC on 'Graphene StressTronics' sponsored by the Communaut\'e Fran\c{c}aise de Belgique, to the European Union through the ETSF e-I3 project (Grant N.$^{\circ}$ 211956), and to the NANOSIM-GRAPHENE project (Projet N.$^{\circ}$ ANR-09-NANO-016-01). P.O. acknowledges Spanish Grants from MICINN (FIS2009-12721-C04-01, CSD2007-00050). Computational resources were provided by the CISM of the Universit\'e catholique de Louvain: all the numerical simulations have been performed on the GREEN and LEMAITRE computers of the CISM.


\begin{thebibliography}{100}
\bibitem{comment1}
 Values of $\xi$ obtained by crossing the solid line with the $G_0$ line in Fig.~\ref{fitSigmaL} are slightly different from the ones obtained in Fig.~\ref{xi}. Numerical inaccuracy originating in the derivative term in Eq.~(\ref{deriveD}) causes the maximum of $\sigma_L$ to be not exactly equal to $\sigma_\text{SC}$.
\end{thebibliography}
\end{document}